\documentclass[a4paper,12pt]{iopart}

\usepackage{times}
\usepackage{cite}
\usepackage{epsfig}
\usepackage{graphicx}

\newcommand{\onlinecite}[1]{\cite{#1}} 
\newcommand{\se}[1]{\emph{#1}} 

\begin{document}


\title[Adsorption of impurities on Si surfaces and
  their influence on etching]{Adsorption of metal impurities on H-terminated Si surfaces and %
their influence on the wet chemical etching of Si %
}
\author{Teemu~Hynninen$^1$, Adam~S~Foster$^1$,
  Miguel~A~Gos\'alvez$^1$, Kazuo~Sato$^2$, Risto~M~Nieminen$^1$}
\address{$^1$
COMP/Laboratory of Physics, 
Helsinki University of Technology, POB 1100, 02015 HUT, Finland.
}
\address{$^2$
Dept. of Micro-Nanosystem Engineering, Nagoya University, 
Nagoya 464-8603, Japan
}
\ead{tjh@fyslab.hut.fi}


\date{\today}

\begin{abstract}
We use first-principles methods to investigate the adsorption of Cu,
Pb, Ag, and Mg onto a H-terminated Si surface. We show that Cu and Pb
can adsorb strongly while Ag and Mg are fairly inert. In addition, two
types of adsorption states are seen to exist for Pb.  We also study
the clustering energetics of Cu and Pb on the surface and find that
while Cu clusters eagerly, Pb may prefer to form only small clusters
of a few atoms. This kind of behavior of impurities is incorporated in
kinetic Monte Carlo simulations of wet etching of Si. The simulation
results agree with experiments supporting the idea that micromasking
by Cu clusters and Pb atoms is the mechanism through which these
impurities affect the etching process.
\end{abstract}

\submitto{New Journal of Physics}
\maketitle

%
%

\section{Introduction}
\label{Introduction}

Anisotropic wet chemical etching of silicon is an important method in
the fabrication of micro-electromechanics systems (MEMS) such as
cantilevers \cite{Saya}, microfluidic channels \cite{Kwon2002}, and
inertial sensors \cite{Schropfer1998}.  However, as structures become
smaller, possible defects and rough features can become detrimental
for the manufactured devices. Therefore it is important to understand
how the features seen on etched surfaces are created.  In principle,
the etching process should rapidly attack protrusions and lead to
atomistically smooth surfaces  \cite{Hines_rev}.  Still,
impurities \cite{Campbell,Tanaka2000_cu,Tanaka2006}, hydrogen
bubbles \cite{Schroder,Campbell,Haiss} or etchant
inhomogeneities \cite{Garcia2004_PRL,Gosalvez2007_zigzag} can disturb
the process and lead to features such as
hillocks \cite{Campbell,Nijdam,Veenendaal2001,Gosalvez2003_NJP},
zig-zags \cite{Shikida2001,Veenendaal2001_zigzag},
pits \cite{Tanaka2006} or step
bunches \cite{Hines_rev,Garcia2004_JPC,Gosalvez2007_zigzag}.  Since
these morphologies are the result of complicated interplay between the
etchant, the surface, and the possible impurities, the problem of rough
surfaces is also theoretically challenging.

Experimentally, it is known that out of the metal impurities commonly
present in etchant solutions, only Cu and Pb affect the etching
process \cite{Tanaka2000_cu,Tanaka2004,Tanaka2006} while the presence
of metals such as Ag, Al, Cr, Zn, Ni, Fe, and Mg has no
effect \cite{Gosalvez2007_rev} (Mg can nullify the effects of Cu if
both are present in the etchant, though \cite{Tanaka2007}). In general,
having Cu in the etchant results in rough surfaces and sometimes also
in low etch rates. For instance, on the (110) and (100) surfaces, Cu
can induce the appearance of hillocks or etch
pits \cite{Tanaka2006}. Pb only affects the etch rates without changing
the surface morphologies. In the case of Cu, this behavior is understood
to be due to adsorption of the impurity particles on the
surface. Adsorption enables the Cu to locally inhibit etching and act
as a pinning agent \cite{Tanaka2006,Gosalvez2007_rev}. The Si surface
is nearly fully terminated by Si--H bonds during wet etching
 \cite{Rappich1994,Allongue_PRL_1996,Hines_rev} and first-principles
calculations have shown that Cu adsorption on such a surface is
energetically favored \cite{Foster2007_CuAds}. Furthermore, Monte Carlo
simulations incorporating adsorbing impurities have demonstrated that
local pinning by impurities and impurity clusters can change the
morphology of etched surfaces \cite{Gosalvez2003_NJP,Hynninen2008}.
However, a comprehensive theoretical understanding of the influence of
other metal impurities does not exist.

In order to form a complete picture of the behavior of metal
impurities in wet etching of Si, we investigate the adsorption of Pb,
Ag, and Mg in addition to Cu. (Note that the adsorption energies of
individual copper atoms cited in this article have been presented
previously in \onlinecite{Foster2007_CuAds}.) While Cu and Pb are
the most active metals according to the existing experimental
knowledge, Ag and Mg are also examined as examples of the other, inert
impurities. We use first-principles calculations to study the
adsorption energetics of these impurities on the hydrogen-terminated
Si surface. For Cu and Pb, we also investigate the tendency for
forming clusters on the surface. Based on the first-principles data,
we carry out kinetic Monte Carlo simulations of etching under the
influence of different types of impurities. By comparing the resulting
surface morphologies and roughnesses as well as etch rates to 
real etched surfaces, we link the theoretical study to experiments.

%
%

\section{Methods}

\subsection{First-principles calculations}

We calculate the adsorption energies of Cu, Pb, Mg and Ag atoms on
H-terminated Si surfaces at various surface sites using the linear
combination of atomic orbitals based \textsc{siesta} code
 \cite{SIESTA1,SIESTA2} implementing the density-functional theory
(DFT) \cite{Hohenberg,Kohn_Sham}.  For exchange-correlation, we use
the PBE gradient corrected functional \cite{pbe}.  Core electrons are
described by norm-conserving pseudopotentials using the
Troullier-Martins parametrization and including scalar relativistic
corrections. The pseudopotentials were generated in electron
configurations [Ne]$3s^23p^2$ for Si, $1s^1$ for H, [Ar]$4s^13d^{10}$
for Cu, [Xe$4f^15d^{10}$]$6s^26p^2$ for Pb, [Ne]$3s^2$ for Mg, and
[Kr]$5s^14d^{10}$ for Ag. Here square brackets denote the core
electron part. Having tested different basis sets, we use double zeta
with a single shell of polarization orbitals for Si, H, and the $5s$
electrons of Ag and triple zeta with two shells of polarization for
Cu, Pb, Mg, and the $4d$ electrons of Ag.  The convergence of system
properties with respect to $k$ points and mesh was also checked. A ($2
\times 2 \times 1$) Monkhorst-Pack $k$-mesh and an energy cutoff of
150 Ry were found sufficient for these systems. The used energy shift
is 25 meV. Basis set superposition errors are handled using
counterpoise corrections \cite{Boys1970_bsse}. The metal impurities
are treated as neutral atoms in the calculations. The impurities
present in an etchant are originally in an ionic form in the solution,
however, the ions are reduced at the surface and adsorb in the neutral
state \cite{Foster2007_CuAds,Tanaka2000_cu,Tanaka2006}.

\begin{figure}[htbp]
\begin{center}
\epsfig{file=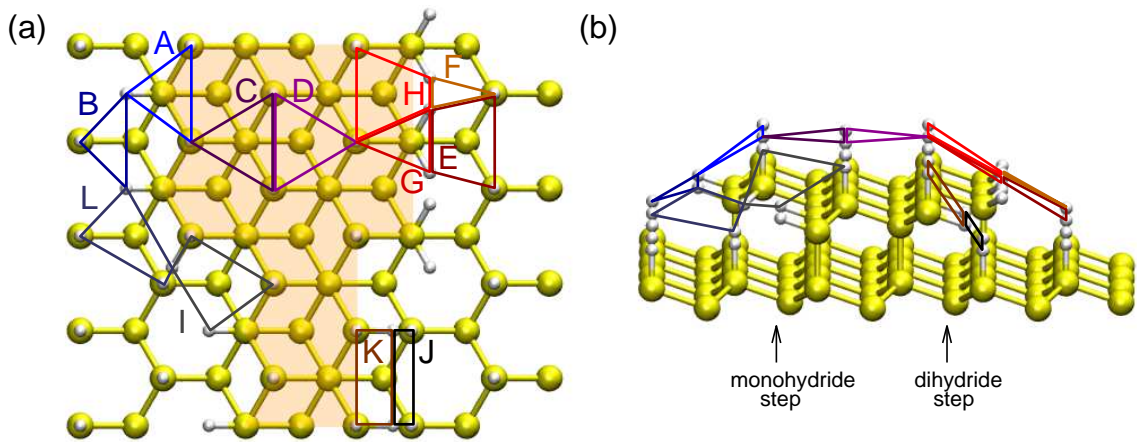,width=15.0cm}
\caption{ The most important adsorption sites on the H-terminated Si
surface found at monohydride steps (\se{A}, \se{B}), (111) terraces (\se{C}, \se{D}),
horizontal dihydride steps (\se{E}--\se{H}), vertical dihydride steps (\se{J}, \se{K}) and
kinks (\se{I}, \se{L}). The large yellow spheres are Si and the small white ones
are H. The surface is shown from (a) top-down and (b) side
perspectives. Two layers of silicon are shown and the upper layer is
shaded in (a) to show the steps better.}
\label{fig:sites}
\end{center}
\end{figure}

The calculations are carried out in supercells of 79--113 atoms
forming two-dimensional Si slabs with both sides terminated by H atoms
in order to saturate dangling bonds. The convergence with respect to
slab thickness was tested and three or four layers of Si atoms were
found to be sufficient. The bottom layer of hydrogen and the lowest
layer of silicon are kept frozen, mimicking bulk silicon. Other atoms
are allowed to relax using a conjugate gradient scheme until atomic
forces become 0.02 eV/\AA\ or less. The surface orientations of the
slabs used in the calculations are (111), (112) and (221). This allows
us to probe the adsorption energies of impurities at a variety of
sites on both ideal (111) terraces as well as on stepped or kinked
surfaces. The most important site types that appear on realistic
surfaces are shown in Figure~\ref{fig:sites} using the naming
conventions of \onlinecite{Foster2007_CuAds, Hynninen2008}. (The
image does not represent a calculation supercell.) Note that two kinds
of steps are shown in the image. The Si in the step on the left share
a bond with one H while on the right there are two H-saturated bonds
per Si. These step types are called mono- and dihydride steps and they
appear on the (221) and (112) surfaces, respectively.

\subsection{Kinetic Monte Carlo simulations}

We model the etching process using an atomistic kinetic Monte Carlo
(KMC) approach. The possible events in the simulation are Si removal,
Cu/Pb adsorption, and Cu/Pb desorption and at each KMC step the
simulator stochastically decides between these according to their
relative occurrence rates. Our implementation
\cite{Gosalvez2002_2,Gosalvez2007_rev} uses the K-level search
algorithm \cite{Blue,Gosalvez2008_KLS} and an event-tree which
combines both Si etching and impurity adsorption and desorption
\cite{Hynninen2008}. The Si are treated as an ideal lattice where the
removal rates for all atoms are determined by their local
neighborhoods, i.e., the numbers of the first and second nearest
neighbor Si atoms on surface and in bulk \cite{Zhou}. The relative
removal rates for the most common sites are given in
Table~\ref{table:rates}. A more thorough classification of the Si atom
types on the surface as categorized by the simulator is given in
\onlinecite{Gosalvez2008_CCA}.  Low etchant concentration is realized
in simulations by setting the etch rate of dihydride steps much higher
than that of monohydride steps. Similarly, high concentration
corresponds to rapidly etching monohydride steps
\cite{Gosalvez2007_rev}.

\begin{table}[htbp!]
\caption{
Relative removal rates for Si atoms per site type in the KMC simulation for low
and high etchant concentrations (Cf. \onlinecite{Gosalvez2007_rev}).
}
\begin{indented}
\item[]\begin{tabular}{lcc}
\br
 & low concentration & high concentration \\
\mr
terrace                  & $10^{-8}$ & $10^{-8}$ \\
monohydride step         & $10^{-6}$ & $10^{-3}$ \\
monohydride kink         & $10^{-4}$ & $10^{-1}$ \\
monohydride double kink  & $10^{-0}$ & $10^{-0}$ \\
dihydride step           & $10^{-2}$ & $10^{-6}$ \\
dihydride kink           & $10^{-1}$ & $10^{-4}$ \\
dihydride double kink    & $10^{-0}$ & $10^{-0}$ \\
\br
\label{table:rates}
\end{tabular}
\end{indented}
\end{table}

The surface is characterized by a constantly updated triangulation
mesh based on the positions of H atoms on the surface. This mesh is
generated as a Delaunay triangulation \cite{Delaunay1934_tri} from a
2D projection of the atomic positions. Each triangle in the mesh then
represents a potential adsorption site for an impurity in the
simulation. The triangles are identified according to their shapes and
unique adsorption and desorption rates are assigned to each triangle
type. The rates are estimated as $\nu \exp(-E/k_{B}T)$, where $E$ is
the activation energy, $\nu$ is a parameter, and $k_B T$ is the
temperature (350~K) multiplied by the Boltzmann constant.  The
adsorption rates of Cu are site-dependent and the used activation
energy values reach from 0.1 to 1.5~eV \cite{Hynninen2008}. The
adsorption rates for Pb are treated site-independently with an
activation energy of 0.3~eV for all triangles (see
Subsection~\ref{subsec:ads}). The etchant is treated as a limited
reservoir of impurities which may adsorb on the surface. Once
adsorbed, an impurity pins the Si atoms connected to the adsorption
site triangle. In this model pinning is handled by lowering the
removal rates of the silicons by a factor of $10^{-3}$. Desorption of
impurities can occur thermally with an activation energy of 1.5~eV or
via underetching, i.e., by removing a Si to which the impurity is
connected \cite{Hynninen2008}.

To simulate impurity clustering in a simplified way, the adsorption
rates on sites (triangles) sharing a vertex with the already
impurity-populated sites are multiplied by an enhancement factor. We
use a factor of $\exp(E/k_{B}T)$, where $E=1.0$~eV for Cu and
$E=0.0$~eV (i.e. no clustering) for Pb, as explained in
Subsection~\ref{subsec:clusters}. For a more detailed description of
the simulation method and especially the treatment of Cu clusters, see
\onlinecite{Hynninen2008} (the ``Interaction enhanced
adsorption'' scheme).

\section{Results}

\subsection{Adsorption of individual impurities}
\label{subsec:ads}

The calculated adsorption energies for the different metal impurities
on various surface sites are given in Table~\ref{table:ae}. Since we
know from experiments that only Cu and Pb affect the etching process, most
effort has been allocated for studying these metals. Two values have
been calculated also for Mg and Ag for comparison.

\begin{table}[htbp!]
\caption{ 
Adsorption energies (in eV) of metal impurities on various surface
sites on the H-terminated Si surface. For Pb, two adsorption states
are distinguished: (A) denotes a weakly adsorbed state and (R) denotes a
strongly bonded state where a Si--H + Pb $\to$ Si--Pb--H reaction has
occurred. An asterisk (*) denotes the state which was found in
structural optimization where the Pb atom was initially placed in the
middle of the site. Cu values are from \onlinecite{Foster2007_CuAds}.
}
\begin{indented}
\item[]\begin{tabular}{cccccc}
\br
site  &    Cu & Pb (A) & Pb (R) &    Mg &    Ag\\
\mr
\se{A} & $-0.55$ & $-0.49$ * & $-1.22$ \phantom{*} & -     & - \\
\se{B} & $-0.89$ & $-0.48$ * & $-1.22$ \phantom{*} & -     & - \\
\se{C} & $-0.77$ & $-0.50$ * & $-1.26$ \phantom{*} & $-0.04$ & $-0.12$ \\
\se{D} & $-1.17$ & $-0.39$ * & $-1.26$ \phantom{*} & -     & - \\
\se{E} & $-1.01$ & $-0.49$ * & $-2.12$ \phantom{*} & -     & - \\
\se{F} & $-1.43$ & -         & $-1.56$ *           & $+0.42$ & $-0.31$ \\
\se{G} & $-0.63$ & $-0.54$ \phantom{*} & $-1.64$ * & -     & - \\
\se{H} & $-1.39$ & -         & $-2.12$ *           & -     & - \\
\se{I} & $-1.34$ & -         & $-1.43$ *           & -     & - \\
\se{J} & $-1.10$ & -         & $-1.46$ * & - & - \\
\se{K} & $-0.50$ & -         & $-1.46$ * & - & - \\
\br
\label{table:ae}
\end{tabular}
\end{indented}
\end{table}

The results for Cu have been previously presented and analyzed in
detail in \onlinecite{Foster2007_CuAds}. In summary, adsorption
of Cu is energetically favorable, with the site-dependent adsorption
energies being approximately proportional to the number of bonds the
Cu atoms can create with the surface H and Si.  The adsorbed Cu atoms
typically sit in between H atoms in the middle of the triangular or
quadrangular sites drawn in Figure~\ref{fig:sites}. The highest
adsorption energy (in absolute value), $-$1.43~eV, is on the \se{F} site found on
vertical dihydride steps. Also the \se{H} and \se{I} sites found on dihydride
steps and kinks, respectively, are favored by Cu.

\begin{figure}[htbp]
\begin{center}
\epsfig{file=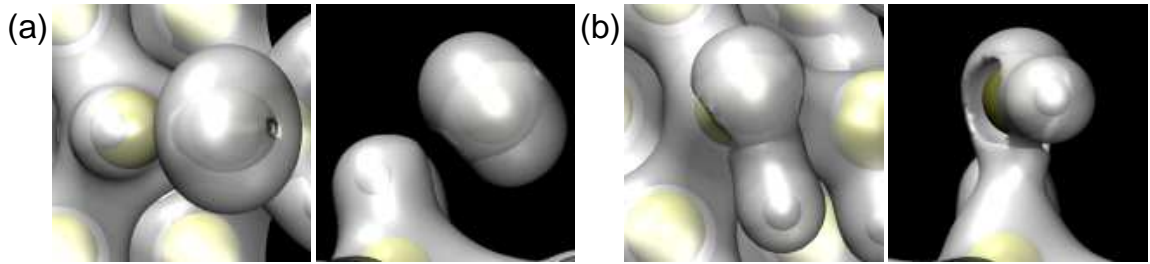,width=15.0cm}
\caption{ Electron densities of the (a) weakly and (b) strongly
adsorbed states of Pb at a \se{G} site on a dihydride step (the step runs
at the right edge of the images) from top-down and side perspectives
drawn at an isosurface of 0.014~e/\AA$^3$.  }
\label{fig:pbstates}
\end{center}
\end{figure}

For Pb, we find two qualitatively different adsorption states. As an
example, electron density plots of these states at the \se{G} site are
shown in Figure~\ref{fig:pbstates}.  Like Cu, Pb can adsorb on the
H-terminated surface. However, the bonding between Pb and H atoms is
weak and the adsorption energies are lower than for
Cu [column ``Pb (A)'' in Table~\ref{table:ae}]. Additionally, Pb does
not exhibit strong site specificity like Cu does --- adsorption
energies are close to $-$0.5~eV regardless of the adsorption site.
The lowest energy found is $-$0.39~eV for the terrace site \se{D}, and the
highest is $-$0.54~eV for the dihydride step site \se{G} shown in
Figure~\ref{fig:pbstates} (a).

The other possibility, shown in
Figure~\ref{fig:pbstates} (b), is that a Pb replaces a hydrogen on the
surface in a reaction Si--H + Pb $\to$ Si--Pb--H. The energy gain in
this reaction is always higher than in plain adsorption of Pb or Cu
[column ``Pb (R)'' in Table~\ref{table:ae}] and the reaction can
happen on any site. The energy change when replacing the H atom on a
monohydride step (corresponds to the shared vertex for sites \se{A} and \se{B}
in Figure~\ref{fig:sites}) is $-$1.22~eV. (The other vertices of the \se{A} and
\se{B} triangles in Figure~\ref{fig:sites} are not equivalent, but reactions
where Hs at these sites are replaced have not been calculated.)  On the (111)
terrace (H shared by sites \se{C} and \se{D}), the energy is $-$1.26~eV.
However, on these sites, merely placing a Pb on the site and letting
the geometry relax results in the weakly adsorbed state. (This is
denoted by asterisks in Table~\ref{table:ae}.) The reacted
states are reached from initial configurations where one hydrogen atom
is slightly pulled away from the surface allowing the Pb to bond
between Si and H. 

On the dihydride step, a few different configurations are found
depending on the initial position from which the Pb atom is allowed to
relax. Placing a Pb on the \se{G} site results in a reaction with the
hydrogen on the upper terrace near the step (H shared by sites \se{D} and
\se{G}) with an energy of $-$1.64~eV. The resulting structure is shown in
Figure~\ref{fig:pbstates} (b). (In this case the weakly adsorbed state
is reached by placing the Pb initially relatively far away from the
surface.)  Starting from the \se{F} or \se{H} sites leads to a reaction with an
H on the horizontal dihydride step (H shared by sites \se{E}--\se{H}), but the
obtained final configurations differ in the orientation of the
Si--Pb--H structure. A configuration where the hydrogen in the Pb--H
pair ends up on the side of the upper terrace (relaxation of Pb
started from site \se{F}) is a local minimum with an energy of
$-$1.55~eV. The opposite case, where the hydrogen is on the side of
the lower terrace (relaxation started from site \se{H}) is the real minimum energy
state for this site with a very low energy of $-$2.12~eV. Placing the
Pb atom initially on the \se{E} site leads to a weakly adsorbed
state. Still, the reacted state with the energy of $-$2.12~eV is
reachable also from the \se{E} site if the corresponding hydrogen atom is
initially pulled away from the surface. Finally, on sites \se{K} and \se{J} a Pb
atom replaces the H on the vertical dihydride step (shared by sites \se{K}
and \se{J}) with an adsorption energy of $-$1.46~eV.

Since the Si--Pb--H states can be reached by the structural
optimization alone on some sites, the energy barrier for the reaction
is likely to be low. To verify this, we calculate the energy barriers
for adsorption and the H replacing reaction on the \se{D} terrace site
using the drag method. Indeed, we find no barrier for a free Pb atom
for reaching the weakly adsorbed state and the barrier between the
weak and strong adsorption states is also low, only about 0.3~eV. This
suggests that it is relatively easy for a Pb atom to adsorb and react
with the surface.

To compare the Cu and Pb results with other metals that do not affect
the etching process, we also calculate the adsorption energies of Mg
and Ag in two test cases. Here, Mg is an example of a metal whose
general properties are very different compared to Cu or Pb while Ag is
interesting since it resembles Cu quite closely (for instance, the
electronic configurations and electronegativities of Cu and Ag are
similar). We calculate their adsorption energies on sites \se{C} and \se{F}
found on (111) terraces and dihydride steps, respectively. The
obtained values are $-$0.04 and $+$0.42~eV for Mg and $-$0.12 and
$-$0.31~eV for Ag, i.e., the energy of an adsorbed Mg is
higher or roughly equal to that of a free atom, and for Ag the energy
gain in adsorption is quite small. This suggests that these metals do
not adsorb on the surface during etching, explaining why they have no
effect on the process.

\begin{figure}[htbp]
\begin{center}
\epsfig{file=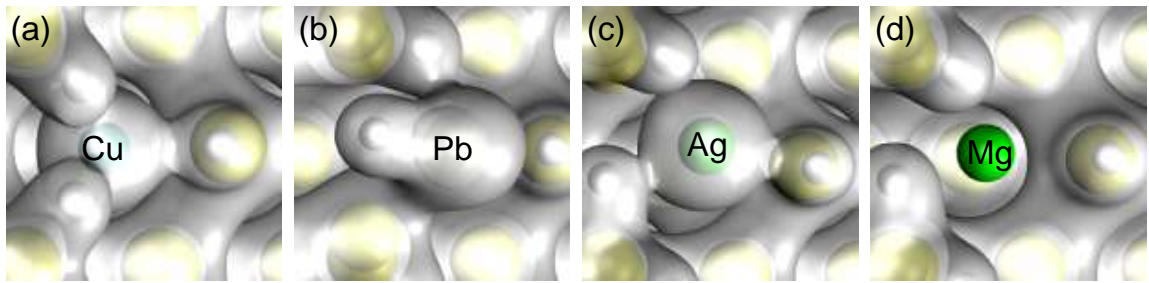,width=15.0cm}
\caption{ Electron densities of (a) Cu, (b) Pb, (c) Ag,
and (d) Mg adsorbed at an \se{F} site on a dihydride step (the step runs at
the left edge of the images). The density isosurfaces are drawn at an
isosurface of 0.014~e/\AA$^3$ for Pb and 0.019~e/\AA$^3$ for the other
atoms. (Only valence electrons are plotted and therefore the density
is lower around Pb than Cu or Ag.) }
\label{fig:fsite}
\end{center}
\end{figure}

The electron densities for all four impurities adsorbed at site \se{F} on a
dihydride step are shown in Figure~\ref{fig:fsite}. Cu, Ag and Mg all
sit between the three surrounding Hs, close to the topmost Si
layer. In the case of Ag and Mg [Figs.~\ref{fig:fsite} (c)
and (d)], the horizontal dihydrides (seen at the left side of the images)
twist, while in the presence of Cu the structure of the step is almost
undisturbed. Also, while Cu clearly forms bonds with the hydrogens, the
electron cloud surrounding Ag only slightly overlaps with the
electrons of the Hs. Mg loses its $s$ electrons completely and
becomes ionic so no bonds are seen (the core electrons are not plotted).
Pb is shown in the Si--Pb--H state, where it has replaced a H that
was bonded with the Si in the lower left corner of the image.

\subsection{Clustering of adsorbed impurities}
\label{subsec:clusters}

\begin{figure*}[htbp!]
\begin{center}
\epsfig{file=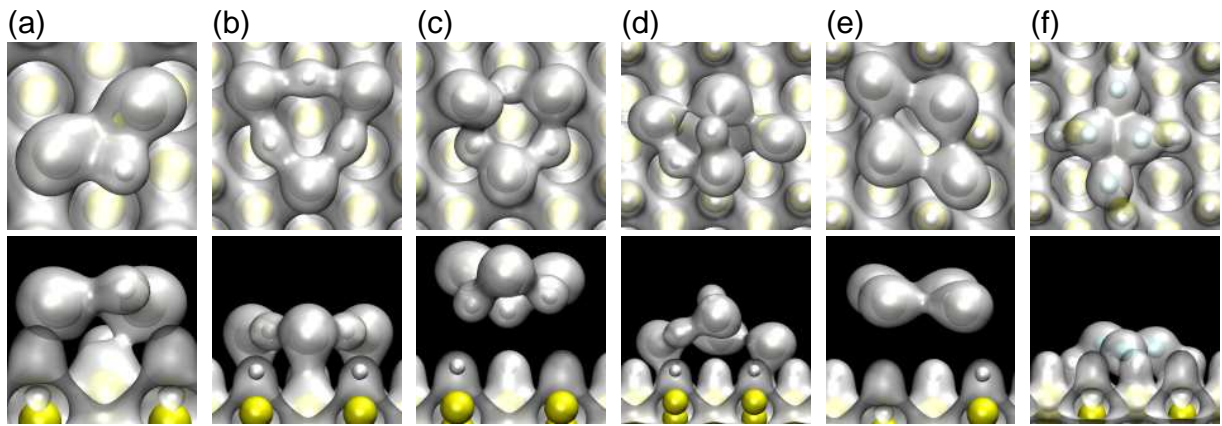,width=16.0cm}
\caption{  Electron densities of (a) a Pb$_2$H dimer,
(b) a Pb$_3$H$_3$ trimer, (c) a Pb$_3$H$_3$ trimer on a H-terminated
surface, (d) a Pb$_4$H$_2$ tetramer, (e) a Pb$_4$ tetramer, and (f) a
Cu$_4$ tetramer shown from top-down and side perspectives. The
densities are drawn at an isosurface of 0.014~e/\AA$^3$ for Pb and
0.019~e/\AA$^3$ for Cu.  }
\label{fig:clusters}
\end{center}
\end{figure*}

Although the calculated adsorption energetics agree with experiments
by suggesting that Cu and Pb can adsorb on the Si surface while other
impurities cannot, they do not explain, for instance, how Cu changes
the surface morphology of the (110) orientation to that covered in
hillocks and Pb only slows down the etch rate \cite{Tanaka2006}. In
\onlinecite{Hynninen2008}, it was argued based on KMC simulations
that Cu impurities should be able to cluster on the surface and that
these clusters are necessary to stabilize the apices of the
hillocks. Therefore, studying the tendencies of Cu and Pb to form
clusters on the surface may be the key to understanding the mechanisms
by which the impurities influence the etching process.  We calculate
the structures and relative energies of Cu and Pb clusters of up to
four impurity atoms on a (111) terrace. As the size of the cluster is
increased, the number of possible stable or metastable structures also
increases. Some of the energetically most favored configurations are shown
in Figure~\ref{fig:clusters}.

Like individual Pb atoms, Pb atoms in clusters can replace Hs on the
surface [see Figs.~\ref{fig:clusters} (a), (b), and (d)]. In this case
the Pb atoms bond strongly with the exposed Si atoms and the removed H
atoms become a part of the cluster.  The other possibility is that the
Pbs cluster with each other and this cluster then weakly adsorbs on
the surface [Figure~\ref{fig:clusters} (e)].  For a Pb dimer, the energy
of a Pb-H-Pb structure shown in Figure~\ref{fig:clusters} (a) is 0.99~eV
below the energy of a weakly adsorbed Pb-Pb structure. (This latter
structure truly is weakly adsorbed: it has an adsorption energy of
only $-$0.14~eV.)  For a Pb trimer, the most stable structure found is
the symmetric Pb$_3$H$_3$ ring shown in Figure~\ref{fig:clusters}
(b). Each Pb in the ring has bonded with a Si by replacing a H and
these Hs then connect the Pb atoms to each other.

We also calculate the binding energy of a Pb$_3$H$_3$ ring on an
undisturbed H-terminated surface [Figure~\ref{fig:clusters} (c)] and
find that there is no interaction with the surface: the adsorption
energy is only $-$0.06~eV. This means that if the Pb atoms reacted
with the H$_2$ released in etching by forming such clusters already in
the etchant, these clusters would not adsorb at all. Therefore, the
hydrogen in the adsorbed clusters should at least initially
come from the Hs removed from the surface by the Pb replacement
reaction.

For tetramers, the only stable symmetric structure found is a metallic
Pb$_4$ cluster [Figure~\ref{fig:clusters} (e)] despite running several
calculations starting from various initial configurations. Even if the
initial configuration is symmetric, the structural optimization
results in the tetramer breaking up. The tetramer structure with the
lowest calculated energy is shown in Figure~\ref{fig:clusters} (d). Two
Hs have been removed from the surface and the left- and rightmost Pb
atoms are bonded with Si. The Hs have moved to the left side of the
cluster while the top- and rightmost Pbs in the image are only weakly
interacting with the other Pb and H atoms.  Interestingly, the energy
of the weakly adsorbed Pb$_4$ [Figure~\ref{fig:clusters} (e)] is only
0.23~eV above the energy of this Pb$_4$H$_2$ tetramer, and all other
calculated tetramers containing strong Si--Pb bonds have an even
higher energy. This suggests that as the number of Pb atoms increases,
the purely metallic clusters become more favorable than the structures
containing hydrogen atoms. On the other hand, since these clusters
adsorb so weakly on the H-terminated surface, they should not
influence the etching process by acting as pinning agents if they do
indeed form.

\begin{figure}[htbp]
\begin{center}
\epsfig{file=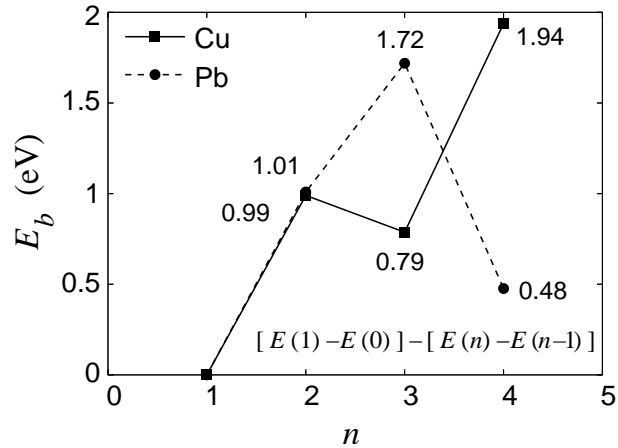,width=8.0cm}
\caption{ 
Binding energies of Cu and Pb in clusters of a few atoms
with respect to the adsorption energy of a single atom.  }
\label{fig:bind}
\end{center}
\end{figure}

Cu atoms on the other hand are eager to form clusters by bonding both
with each other and with the H atoms on the surface, as shown for a
tetramer in Figure~\ref{fig:clusters} (f). This suggests that Cu should
have no difficulties in forming quite large clusters on the
surface. As a measure of the clustering tendency, we plot the binding
energies $E_b$ of impurities in clusters as a function of cluster size
$n$ in Figure~\ref{fig:bind}.  The binding energy is calculated as
$E_b(n) = [E(1)-E(0)] - [E(n)-E(n-1)]$, where $E(n)$ is the energy of
the energetically most favorable system containing the Si(111) slab
and an $n$-mer.  In other words, $E_b$ measures the energy gain from
joining an impurity to a cluster of $n-1$ atoms to form the $n$-mer
against the energy associated with the adsorption of the impurity
alone as a monomer. If $E_b$ is positive, clustering is favored.

We see that the $E_b$ graphs for Cu and Pb are qualitatively
different. $E_b(1)$ is 0~eV by definition, and also the dimer binding
energy $E_b(2)$ is the same, 1.0~eV, for both metals. (Cf. with 0.9~eV
for Cu in \onlinecite{Hynninen2008} calculated by comparing a
system with a dimer to one with two monomers present.) Since the
symmetric Pb$_3$H$_3$ ring is very stable, $E_b(3)$ is 1.72~eV for Pb
whereas it is only 0.79~eV for Cu, i.e., even slightly lower than
$E_b(2)$.  Most importantly though, $E_b(4)$ for Pb drops dramatically
to only 0.48~eV, more than one eV below the trimer binding energy. Cu
behaves in a completely opposite way as $E_b(4)$ climbs to 1.94~eV.
Despite the small decrease at $n=3$, the binding energies for Cu seem
to grow when cluster size increases. (The average binding energy per
Cu atom, $[E(0)-E(n)]/n$, does always increase as a function of $n$.)
This indicates that it is energetically favorable to create quite
large clusters. The binding energies of Pb in dimers and trimers are
also considerable suggesting that also Pb may form clusters. However,
the slump seen in tetramer binding energies means that although
joining a Pb in a trimer in order to form a tetramer is energetically
slightly better than having a monomer and a trimer [$E_b(4) > 0$], it
is more favorable to join the Pb with another monomer or a dimer
[$E_b(4) < E_b(3)$]. The optimal Pb configuration may in fact consist
of many small clusters --- preferably the ring-shaped trimers --- and
of no large complexes. (The average binding energy per Pb atom is
lower in tetramers than in trimers.)

\subsection{Influence of impurities on etching}

\begin{figure*}[htbp!]
\begin{center}
\epsfig{file=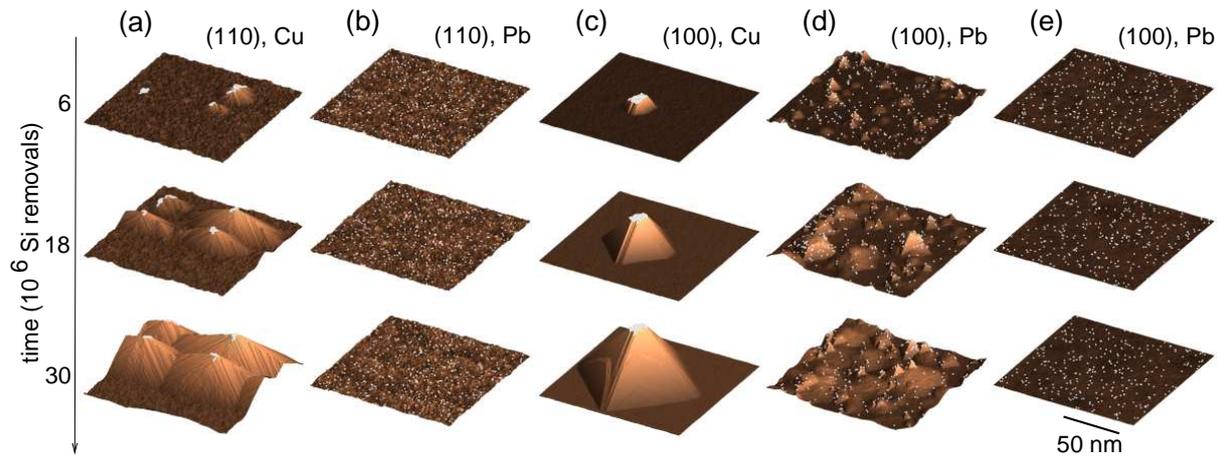,width=16.0cm}
\caption{  Evolution of the (110) [(a)--(b)] and (100)
[(c)--(e)] surface morphologies in KMC simulated etching under the
influence of Cu (clustering impurities) [(a) and (c)] and Pb
(individually adsorbing impurities) [(b), (d), and (e)]. In (e), the
removal rates of monohydride step and kink sites are a factor of ten
higher than in (d).  The size of the simulated systems is 120 nm
$\times$ 120 nm. Impurities are shown as white dots.  }
\label{fig:evol}
\end{center}
\end{figure*}

In order to quantify how clustering of impurities affects the etching
process, we simulate the etching of (110) and (100) surfaces. Cu is
simulated as an impurity that forms clusters. The site specific
adsorption rates for Cu are set to be the same as in
\onlinecite{Hynninen2008}, following the first-principles
adsorption activation energy hierarchy. Pb on the other hand is
treated as an impurity that does not cluster. Although our
first-principles results suggest that Pb may form clusters of a few
atoms, the pinning effect of such small particles should not be much
stronger than that of individual atoms and so we ignore the formation
of Pb clusters altogether. We use a constant adsorption rate for Pb
everywhere since the adsorption energy barriers should be low
according to the first-principles results.  Figure~\ref{fig:evol}
shows the time evolution of the (110) surface in simulated etching at
high etchant concentration and the (100) surface at low concentration
(see Table~\ref{table:rates}). We study these specific surfaces and
conditions since trapezoidal and pyramidal hillocks are seen in the
corresponding experiments \cite{Gosalvez2007_rev}.

In Figure~\ref{fig:evol} (a), Cu clusters are seen on (110) and
trapezoidal hillocks develop underneath. Since the lifetime of the
clusters and also hillocks is long, the limited size of the simulated
system is reached and the hillocks start merging. Conversely, there is
no mechanism that drives cluster formation in the Pb
simulations. Since Pb adsorption is uniform, the impurities are
scattered randomly on the surface, as shown in Figure~\ref{fig:evol}
(b). The surface morphology remains unchanged as etching proceeds.

In Figure~\ref{fig:evol} (c), the (100) surface is seen when etched
under the influence of Cu. A Cu cluster is protecting the apex of a
pyramidal hillock. Again, the hillock is stable enough to grow until
it reaches the size of the simulation supercell. The effect of Pb on
simulated (100) depends on the etchant concentration, however.  If the
etching of dihydride steps is fast enough with respect to the
monohydrides (as in Table~\ref{table:rates}), small hillocks are seen
[Figure~\ref{fig:evol} (d)]. Since the ridges of the hillocks consist of
Si atoms similar to those on monohydride steps and the facets are
formed by (111) terraces, the hillocks become very stable in these
conditions and even individual impurities can stabilize them. Still,
the impurities are removed from the surface quite quickly due to
underetching and the lifetime of these hillocks is short. The steady
state morphology of the surface, where the surface is textured with
the small hillocks and their remains, is reached quickly. If the
relative stability of monohydrides is weakened slightly (the removal
rates of Si atoms on monohydride step and kink sites are increased by
a factor of 10 from the previous situation), hillocks remain quite
stable and grow underneath impurity clusters [Figure~\ref{fig:evol} (c)]
but not individual impurities [Figure~\ref{fig:evol} (e)] due to faster
underetching.

\begin{figure}[htbp!]
\begin{center}
\epsfig{file=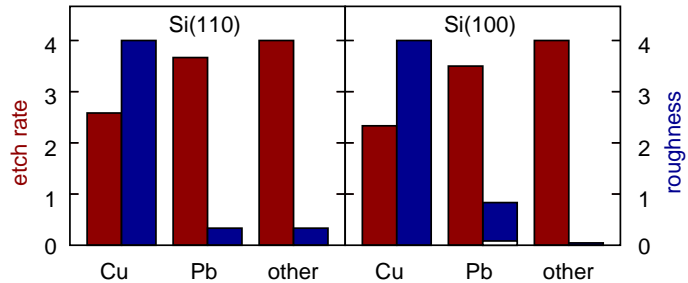,width=9.0cm}
\caption{ Average etch rate (left, red bars) and roughness (right,
blue bars) of simulated 120 nm $\times$ 120 nm (110) and (100)
surfaces with Cu, Pb, and other impurities in arbitrary units. The
(110) results are for high and (100) for low etchant
concentrations. For Pb on (100), the blue and white bars correspond to
the morphologies in Figs.~\ref{fig:evol} (d) and (e), respectively.
}
\label{fig:rr}
\end{center}
\end{figure}

Since the simulated Cu clusters are so stable that they enable
hillocks to grow larger than feasible simulation cell sizes, it is not
possible to reliably measure the impurity effect on etch rate and
surface roughness from these simulations. In addition, the applied
simulation method does not include the necessary effects to simulate
e.g. the formation of deep etch pits which also affect these
properties. Still, we make a qualitative comparison between the
different cases presented above by calculating the etch rates and
roughnesses (standard deviation of height). The properties are
calculated for 120~nm$\times$120~nm systems with 435 impurity atoms
present [3~impurities/(10~nm)$^2$] after 30 million Si removal events
(etch rate is measured for the final 6 million events)
and averaged over five simulations. The obtained values are plotted in
Figure~\ref{fig:rr}. In addition to simulating etching with Cu and Pb,
we also investigate the case where no impurities are adsorbing on the
surface. Note that while the surfaces containing Pb or other
non-adsorbing impurities
reach a steady state during the simulation, the systems with Cu do not.

For (110), we find that Cu greatly increases surface roughness
compared to when no impurities are present since it induces hillock
growth. Cu also lowers the etch rate considerably. Pb on the other
hand affects neither the surface morphology nor roughness. Pb does
lower the etch rate, though. In fact, the simulated etch rate
decreases further when the impurity concentration is increased (not
shown). The relative etch rates of (100) with the different impurities
are very close to those obtained for (110): Cu and Pb decrease the
etch rate by about one third and one eighth, respectively. 
The Pb roughness
shown as a blue bar in Figure~\ref{fig:rr} corresponds to the
morphology of Figure~\ref{fig:evol} (d) with small hillocks. The
roughness of this surface is twenty times the roughness of the systems
with no adsorbing impurities, although still clearly below the
roughness obtained with Cu. [The undisturbed simulated (100) is almost
atomistically smooth.]  For the hillockless morphology of
Figure~\ref{fig:evol} (e) (white bar in Figure~\ref{fig:rr}),
roughness is almost as low as for the system with no
impurities. Similar change in etching conditions has only a marginal
effect for the other impurities and also the average etch rate with Pb
remains virtually unchanged between the systems shown in
Figs.~\ref{fig:evol}~(d) and (e).

\section{Discussion}

Tanaka et al. \cite{Tanaka2006} explained the experimentally observed
effects of Cu and Pb on the etching process using an argument based on
the oxidation-reduction potentials at high pH: Since the potentials of
Cu and Pb, $-$0.40 and $-$0.91~V, respectively, are roughly the same
or higher than that of hydrogen, $-$0.85~V, the H$_2$ released in the
etching process reduces the ionic impurities to a neutral state
allowing them to adsorb in metallic form. Most of the impurities have
low potentials and thus they remain as ions that do not adsorb.  On
the other hand, the potential of Ag is high, $+$0.40~V, and it should
also be in the metallic state according to this argument. However,
since our calculations show that the adsorption energy of metallic Ag
on the H-terminated Si surface is close to zero, we argue that Ag
does not adsorb even in the neutral state and thus it does not affect
the etching reactions.

The tendency to react with the surface seen with Pb but not with the
other impurities can be understood by examining electronegativities.
Pb is more electronegative than H, the Pauling values being 2.3 and
2.2, respectively, and so Pb seeks additional electrons more eagerly
than H. Therefore, instead of only bonding with the surface hydrogens,
it is energetically favorable for Pb to bond directly with the
silicons in a Si--Pb--H structure where a H is replaced by the
Pb. Since Cu and Ag have lower electronegativities, 1.9 both (same as
Si), they cannot replace Hs like Pb can. The electronegativity of Mg
is so low (1.3) that the atom would give away its outer electrons to
the Si surface if it was adsorbing in the neutral state, as seen in
Figure~\ref{fig:fsite} (d).  (Although, according to the argument based
on oxidation-reduction potentials, Mg should not be adsorbing in the
neutral state in the first place.)

The adsorption behavior of Cu and Pb are qualitatively different. Cu
adsorption energies are strongly site-specific while for Pb the
adsorption energy depends mostly on the type of adsorption state
rather than the adsorption site.  However, the most important
qualitative difference between Cu and Pb, when etching is considered,
is the clustering tendency of Cu. Although the first-principles
calculations do not prove that Pb cannot cluster on the Si surface,
they suggest that Pb forms only small complexes. The energetically
most favorable configuration found for Pb is a trimer.  If the number
of Pb atoms is increased, a metallic cluster consisting purely of Pb
may become the lowest energy configuration. However, according to the
first-principles calculations, such Pb clusters interact only weakly
with the H-terminated surface (unlike Cu clusters). In this regard the
two different adsorption states of Pb play an important part in
separating Pb and Cu, since Pb needs to replace the hydrogens on the
surface in order to adsorb strongly --- something the metallic cluster
does not do.
 
The surface morphology features obtained in KMC simulations based on
the picture of impurities that either do or do not form clusters agree
with experimentally seen morphologies and support the idea. More
importantly, the estimated etch rate and surface roughness trends for
(110) agree with the measured values in \onlinecite{Tanaka2000_cu}. In
experiments, Cu was seen to increase surface roughness by a factor of
four by inducing the growth of hillocks.  Cu also lowered the
experimentally measured etch rate by some tens of percents depending
on the Cu concentration.  These effects are also seen in the
simulations.  (The increase in simulated roughness is by a factor of
eight instead of four, though, and it would be even greater if the
simulated systems were larger.  However, the surfaces seen in
experiments are much rougher than what can be realized in the small
simulated systems in any case.)  The increased roughness is obviously
due to hillocks but, in fact, so is the decreased etch rate: The
rapidly etching (110) floor gets replaced by the approximately \{311\}
oriented, slowly etching hillock facets and this slows down the
average etch rate of the whole surface
\cite{Tanaka2000_cu,Gosalvez2007_rev}. Pb on the other hand does not
affect the surface roughness but lowers the etch rate, again in agreement
with the experiments. This is due to the impurities adsorbing on the
surface and slowing the etching process locally. Since the Pbs
constantly desorb and adsorb, the morphology is left unchanged and
effectively the etch rate of the whole surface is
decreased. Increasing the number of Pb atoms in the simulation
decreases the average etch rate even further. This is true also in the
experiments up to a concentration of 100 ppb \cite{Tanaka2000_cu}.

We are not aware of experiments where (100) is etched at low etchant
concentration while Cu and Pb concentrations are manipulated. [Such
experiments exist for (100) at high concentration, though, but in such
conditions the presence of Cu impurities leads to the appearance of
round pits instead of hillocks \cite{Tanaka2006}.]  Still, it is
experimentally known that pyramidal hillocks are common in these
etching conditions
\cite{Gosalvez2007_rev,Campbell,Nijdam,Veenendaal2001}, although,
their appearance may often be induced by other means than a high
concentration of metal impurities, e.g., by hydrogen bubbles
\cite{Campbell,Schroder,Haiss}. The KMC simulations, where these
additional factors are not present, predict that large pyramidal
hillocks can be stabilized by Cu clusters just like the trapezoidal
hillocks on (110).  In addition, the presence of Pb impurities may
also be sufficient for inducing the growth of small hillocks. The
stability of these small hillocks depends quite delicately on the
etching conditions, though, and it is possible that they do not appear
in real etching.

\section{Conclusions}

We have demonstrated using first-principles calculations that Cu and
Pb adsorb on the hydrogen-terminated Si surface while other common
metal impurities, represented by Mg and Ag, do not. This explains why
only Cu and Pb have been observed to affect the wet etching process of
Si. The adsorption energies of Cu depend strongly on the adsorption
site whereas Pb adsorption is quite homogeneous.  Furthermore,
calculated energies of clusters of up to four metal atoms on the
surface indicate that it is energetically favorable for Cu to form
large clusters, while for Pb the optimal size of a strongly adsorbed
cluster is three Pb atoms.  Clusters of four Pb atoms
containing H tend to break suggesting that large Pb clusters adopt a
metallic state. However, these metallic clusters interact very weakly
with the H-terminated Si surface and so only the small Pb clusters may
act as micromasks during etching.  This picture is supported by
kinetic Monte Carlo simulations of wet etching of Si in which clusters
of adsorbed impurities lead to rough surfaces and decreased etch rates
(compared to etching in the absence of impurities) while individually
adsorbed impurities lower the etch rate but do not affect the surface
roughness.  These two types of behavior correspond to, and explain,
the experimentally observed effects of Cu and Pb impurities,
respectively.

\ack We acknowledge the generous computing resources from the Center
for Scientific Computing, Helsinki, Finland. We thank J. von Boehm for
valuable discussions. This research has been supported by the Academy
of Finland through its Centers of Excellence Program (2006-2011) and
Project SA 122603, by the Japanese MEXT 21st Center of Excellence
Program ``Micro- and Nano-Mechatronics for Information-Based Society''
and the JSPS-Bilateral Program with the Academy of Finland. Atomic
images in this paper were rendered using the VMD visualization
software.

\vspace*{1cm}
\bibliographystyle{unsrt}
\bibliography{thesisrefs}

\end{document}